\documentclass[12pt,preprint]{aastex}

\def\MH{\ensuremath{M_\mathrm{h}}} \def\RH{\ensuremath{R_\mathrm{h}}}

\def\NHH{\ensuremath{N_\mathrm{H_2,h}}}

\def\NHHDH{\ensuremath{N_\mathrm{H_2D^+,h}}}
\def\xHHD{\ensuremath{x_\mathrm{H_2D^+}}}
\def\nHH{\ensuremath{n_\mathrm{H_2}}} 
\def\agr{\ensuremath{a_\mathrm{gr}}} 
\def\zcr{\ensuremath{\xi_\mathrm{cr}}}

\usepackage{natbib}
\usepackage{graphicx}
\usepackage{amsmath}

\begin{document}

%%%%%%%%%%%%%%%%%%%%%%%%%%%%%%%%%%%%%%%%%%%%%%%%%
\title{H$_2$D$^+$: a light on baryonic dark matter?}

\author{
C. Ceccarelli\footnote{Laboratoire d'Astrophysique, Observatoire de Grenoble -
 BP 53, F-38041 Grenoble cedex 09, France : 
 Cecilia.Ceccarelli@obs.ujf-grenoble.fr}  and
C. Dominik\footnote{Astronomical Inst. ``Anton Pannekoek'', Univ. of
 Amsterdam, Kruislaan 403, NL-1098SJ Amsterdam : dominik@science.uva.nl}
}

%\date{Received ... /Accepted ....}

\begin{abstract}
  It has been suggested that the dark halos of galaxies are
  constituted by cloudlets of cold ($\leq 10$ K) H$_2$ and dense
  ($\geq 10^7$ cm$^{-3}$) molecular gas.  Such gas is extremely
  difficult to detect, because the classical tracers of molecular gas,
  CO and/or dust grains, have very low abundances and their emission
  is exceedingly weak.  For this reason, the cloudlet hypothesis
  remains so far substantially unproven. In this Letter we propose a
  new method to probe the presence of cold H$_2$ clouds in galactic
  halos: the ground transition of ortho-H$_2$D$^+$ at 372 GHz.  We
  discuss why the H$_2$D$^+$ is abundant under the physical conditions
  appropriate for the cloudlets, and present a chemical model that
  predicts the H$_2$D$^+$ abundance as function of four key
  parameters: gas density and metallicity, cosmic ray ionization rate
  and dust grain size.  We conclude that current ground-based
  instruments might detect the ortho-H$_2$D$^+$ line emitted by the
  cloudlets halo, and prove, therefore, the existence of large
  quantities of dark baryonic matter around galaxies.
\end{abstract}

\def\fdep{\ensuremath{f_\mathrm{dep}}}
\def\kf{\ensuremath{k_\mathrm{f}}}

\keywords{Cosmology - Dark matter }

%\maketitle
\setcounter{footnote}{0}

%%%%%%%%%%%%%%%%%%%%%%%%%%%%%%%%%%%%%%%%%%%%%%%%%%%%%%%%%%%%%%%%%%%%%
\section{Introduction}

One of the most astounding results of modern cosmology is the fact
that we currently firmly understand the nature of but a tiny fraction
of the matter content of the universe: the ``luminous matter''
constituting the stars and gas in galaxies. This component amounts to
only $\sim$0.3\% of the mass \citep{2004ApJ...616..643F}.  The largest
fraction of cosmic matter belongs to the so--called ``dark sector''
(95.4\%), split into dark energy (72\%) and dark matter (23\%).  In
total, $\sim$4.5\% of the universe matter is present in the form of
baryonic matter.  However, stars account for only $\sim$5\% of it.  A
significant fraction of the remaining 95\% is thought to constitute
the warm/hot intergalactic gas traced by the Ly$\alpha$ forest
\citep{2002ApJ...564..631S,2004ApJS..152...29P,2006A&A...445..827R}
and/or X-rays
\citep{1999ApJ...514....1C,2005Natur.433..495N,2006astro.ph..1620W}.
However, the numbers are uncertain and up to 50\% of the baryons at
low redshift may still be unaccounted for.

In galaxies, the evidence that most of the mass is \emph{not}
contained in the stellar or any other observed component stems from
the study of rotation curves, derived from stellar light and HI gas
emission \citep{1962AJ.....67..491R,1998A&AS..127..117P}, which
indicate the presence of a large invisible mass surrounding the
galaxies.
The nature of these dark matter halos has long baffled astronomers. It
cannot be {\it diffuse} baryonic matter, because both neutral or
molecular gas would be detected in one way or another.  For this
reason, it has been proposed that it may be non-baryonic in nature.
However, the amount of dark mass in galaxies could be as large as the
amount needed in the form of baryons in order to fulfill the
constraints from Big Bang nucleosysthesis \citep{2003ApJS..149....1K}.
Besides, there are other reasons to assume that the dark matter in
galaxies may be mostly baryonic \citep{1994A&A...285...79P,
1994A&A...285...94P, 1996ApJ...472...34G}.  An invisible baryonic form
would be condensed objects, like brown dwarfs or massive compact halo
objects (MACHOS), but they both have now been ruled out as significant
contributors \citep{2001ApJ...550L.169A}.  Another possibility is that
baryonic dark matter is present in the form of cold ($\leq 10$K),
%dense ($\geq 10^6$ cm$^{-3}$)
molecular clouds, also called ``cloudlets''
\citep{1994A&A...285...79P,1994A&A...285...94P}.  There is
considerable discussion in the literature about the allowed range of
parameters for such clouds, and the values obtained depend on the
particular model and physical constraints used.
\citet{1996ApJ...472...34G} consider both the collisional cross
section of clouds and the thermal stability against collapse and
conclude that the clouds should be colder than 10 K, have masses
around 1 $M_\odot$ and column densities above $1\times10^{23}$
cm$^{-2}$.  \citet{1998ApJ...498L.125W} suggest that such clouds can
be responsible for radio scintillations observed toward quasars and
derive that the clouds should be a few AU in size and have masses
below 10$^{-3}$ $M_\odot$. \citet{1999ApJ...527L.109W} consider the
thermal stability of such clouds against evaporation and find a lower
limit for the mass of $\sim 10^{-6}$ $M_\odot$ to keep the clouds
stable in the current epoch (low redshift).  All these estimates
involve uncertain parameters like the cosmic ray ionization rate in
the halo.  \citet{2001ApJ...547..207R} use constrains from
micro-lensing studies to derive a lower limit of the cloud masses of
$\sim 10^{-5} M_{\odot}$ and conclude that models with the Walker and
Wardle parameters of 10 $^{-3}$ $M_{\odot}$ and about 10 AU radii are
still acceptable within the current observational limits.
\citet{1999A&A...350L...9K} and \citet{2004ApJ...610..868O} suggest
small halo cloudlets of 6AU and masses of 10$^{-3}$ $M_{\odot}$ as the
explanation for the unidentified EGRET sources. It is, however, not
clear if these clouds exist in a stable form, or if they are simply
the highest density component of a fractal distribution of cold
molecular gas. In summary, several authors have invoked the
possibility that a substantial amount of baryonic mass is in cold
cloudlets, but the uncertainty in the involved physics is so large
that so far no consensus exists on the mass and size of the presumed
cloudlets.  \citet{2004ApJ...606L..13H} tried to probe the cloudlets
in our own galaxy by means of high angular resolution CO observations.
He indeed detected clumps of molecular material at 7-18 K, but it is
unclear what their masses, sizes and densities are, given the
uncertainty on the distance and CO abundance (see \S 2).  In addition,
\citet{2001MNRAS.323..147L} discusses the possibility that some faint
SCUBA sources are in fact local very low temperature dust clouds, with
characteristics similar to the halo cloudlets.
  
In this Letter, we assume that the galactic dark halos are constituted
of such cloudlets, without further discussing the detailed structure
and stability of such objects.  Instead, we propose a new method to
test this hypothesis: observations of the ground state transition of
the ortho-H$_2$D$^+$.

%%%%%%%%%%%%%%%%%%%%%%%%%%%%%%%%%%%%%%%%%%%%%%%%%%%%%%%%%%
\section{Why H$_2$D$^+$}

Low metallicity cold molecular gas is intrinsically difficult to
detect.  The main constituent of cold gas, H$_2$, does not have a
permanent dipole moment and its ground transition can only be excited
in warm ($>100$K) gas. The most sensitive probes of ``standard''
molecular clouds in the galactic disk are dust continuum emission and
CO millimeter line emission.  However, if the gas is primordial
(i.e. not enriched with the products of stellar nucleosynthesis) or
very low in metallicity, the dust abundance is zero or very low and
its emission undetectable.  The same argument holds for CO.  Even if
the metallicity is non-zero and a small amount of CO and dust were
present, the CO molecules would quickly freeze-out onto dust grains
and in this way still escape detection.
Similar chemical conditions can actually be found in particularly cold
objects of our Galaxy: prestellar cores and protoplanetary disks,
which contain regions so cold and dense, that CO and other
heavy-element bearing molecules freeze out onto the cold grains.
While galactic matter is still traceable through dust emission, the
severe depletion of CO from the gas phase causes these regions to
become ``invisible'' in terms of molecular emission, just like the
presumed cloudlets in the dark galactic halos.  However, recent
breakthroughs in the observation and theoretical understanding of
\textit{molecular deuteration} have opened new ways to probe cold and
CO depleted regions: the ground transition of H$_2$D$^+$ at 372 GHz,
which is indeed a very specific signature of extremely cold and dense
regions.  As a matter of fact, the detection of this line has been
sought after for decades in galactic giant molecular clouds (whose
mass is $\geq 10^4$ M$_\odot$) without success
\citep{1992A&A...261L..13V}, whereas it is easily detected in the
small ($\leq 1$ M$_\odot$) but cold and dense pre-stellar cores
\citep{2003A&A...403L..37C} and/or proto-planetary disks
\citep{2004ApJ...607L..51C}.  The initially unexpected presence of
deuterated molecules in large abundances \citep{2004dimg.conf..473C}
has now been explained by a new class of chemical models
\citep{2003ApJ...591L..41R}.  It is precisely the disappearance of the
CO from the gas phase that is causing a dramatic enhancement of
H$_2$D$^+$, HD$_2^+$ and D$_3^+$ with respect to H$_3^+$ (for details
see \S~\ref{sec:model}).
Excitingly, H$_2$D$^+$ and HD$_2^+$ \emph{do} have ground transitions
observable with ground-based sub-millimeter telescopes
\citep{1999ApJ...521L..67S,2004ApJ...606L.127V}. These transitions
can be used to probe the presence of the hypothetical
cloudlets forming the baryonic dark matter.

%%%%%%%%%%%%%%%%%%%%%%%%%%%%%%%%%%%%%%%%%%%%%%%%%%%%%%%%%%%%%%%%%%
\section{Model Description}
\label{sec:model}

The chemical model we are using is largely identical with the one
described by \cite{2005A&A...440..583C}, which is based upon work by
\citet{2003ApJ...591L..41R} and \citet{2004A&A...418.1035W}.  We
summarize here the basic assumptions entering into this model, and the
modifications we added to meet the requirement of cold gas clouds in
galactic halos.
In cold and dense gas, the most abundant molecule is always H$_2$.  HD
is present as well, and generally is the main reservoir of deuterium,
with an abundance of $3\times10^{-5}$ relative to H$_2$
\citep{2003SSRv..106...49L}.  At low temperature, chemistry is
generally driven by ion-neutral reactions, and therefore, by the process
of ionization due to cosmic rays.  Cosmic rays ionize
H$_2$ and quickly lead to the formation of H$_3^+$, which is the main
charge carrier in the gas.  H$_3^+$ is destroyed by reactions with
grains, heavy element bearing molecules like CO and N$_2$ and, at a
very low rate, through direct recombination with electrons.  It also
reacts with HD, and this is the starting point of deuterium enrichment
chemistry whose overwhelming importance has only recently been
recognized.  H$_3^+$ reacts with HD to form H$_2$D$^+$.
\begin{equation}\label{h3+}
\mathrm{H_3^+ + HD \rightarrow H_2D^+ + H_2}
\end{equation}
The reaction is endothermic, i.e. the reverse reaction H$_2$D$^+$ +
H$_2$ $\to$ H$_3^+$ + HD is practically forbidden at temperatures
below $\sim30$ K.  Therefore, a deuterium atom captured in this way
into an H$_3^+$ isotopologue remains there until it is returned by a
reaction neutralizing the ion.  The relevant reactions are those with
dust grains, CO and N$_2$ molecules, and direct recombinations with
electrons.  As in cold and/or metal-poor environments the abundance of
those species is low, H$_2$D$^+$ tends to include the \emph{entire
positive charge}.  Of course, H$_2$D$^+$ can itself react with HD to
form HD$_2^+$, and a further step leads to D$_3^+$.

\noindent
Therefore, in cold gas, the positive charge is distributed between
H$_3^+$, H$_2$D$^+$, HD$_2^+$, and D$_3^+$
\citep{2003ApJ...591L..41R,2004A&A...418.1035W}.  It turns out that,
at extremely low metallicities, this no longer holds. Under ``normal''
galactic conditions, H$^+$ is quickly destroyed by recombination on
dust grain surfaces.  At the lower grain abundances characteristic of
low metallicity conditions, the only destruction channel for H$^+$ is
direct recombination with electrons, a very inefficient reaction.
Therefore, H$^+$ can dominate the positive charge. In the case where
H$^+$ dominates, the ionization balance can be calculated ignoring the
presence of H$_3^+$ and solving the equation
\begin{equation}\label{eq:1}
x_\mathrm{H^+}^2 n_\mathrm{H_2} k_\mathrm{pe} + 2[\mathrm{D}] x_\mathrm{H^+}^2 n_\mathrm{H_2} k_\mathrm{pHD}  - \zcr = 0
\end{equation}
where $x_{H^+}$ is the H$^+$ abundance with respect to H$_2$,
$n_\mathrm{H_2}$ is the H$_2$ density, [D] is the elemental deuterium
abundance, \zcr is the cosmic ray ionization rate, $k_\mathrm{pe}$ and
$k_{pHD}$ are the reaction rates of H$^+$ recombination with electrons
($3.6\times10^{-12} (T_\mathrm{gas}/300K)^{-3/4}$ cm$^3$ s$^{-1}$;
\cite{1997A&AS..121..139M}) and HD ($2.0\times10^{-9}
\rm{exp}[-464/T_\mathrm{gas}]$ cm$^3$ s$^{-1}$) respectively.  In the
case where H$_3^+$ and its deuterated isotopologues dominate the
positive charge, the electron abundance is found by solving Eq. 13 of
\cite{2005A&A...440..583C}.  
The low metallicity in primordial gas also affects directly the
abundances of CO, N$_2$ and dust grains.  In the present model, the
abundance of the grains $x_\mathrm{gr}$ is scaled with a factor
$Z/Z_\mathrm{sun}$ denoting the metallicity of the material relative
to solar abundances.  In the same way, we scale the abundances of CO
and N$_2$ linearly with the metallicity.

%%%%%%%%%%%%%%%%%%%%%%%%%%%%%%%%%%%%%%%%%%%%%%%%%%%%%%%%%%%%%%%%%%%%%%%%%%%%
\section{Discussion and Conclusions}

\begin{figure*}[htbp]
\centerline{\includegraphics[width=15.0cm]{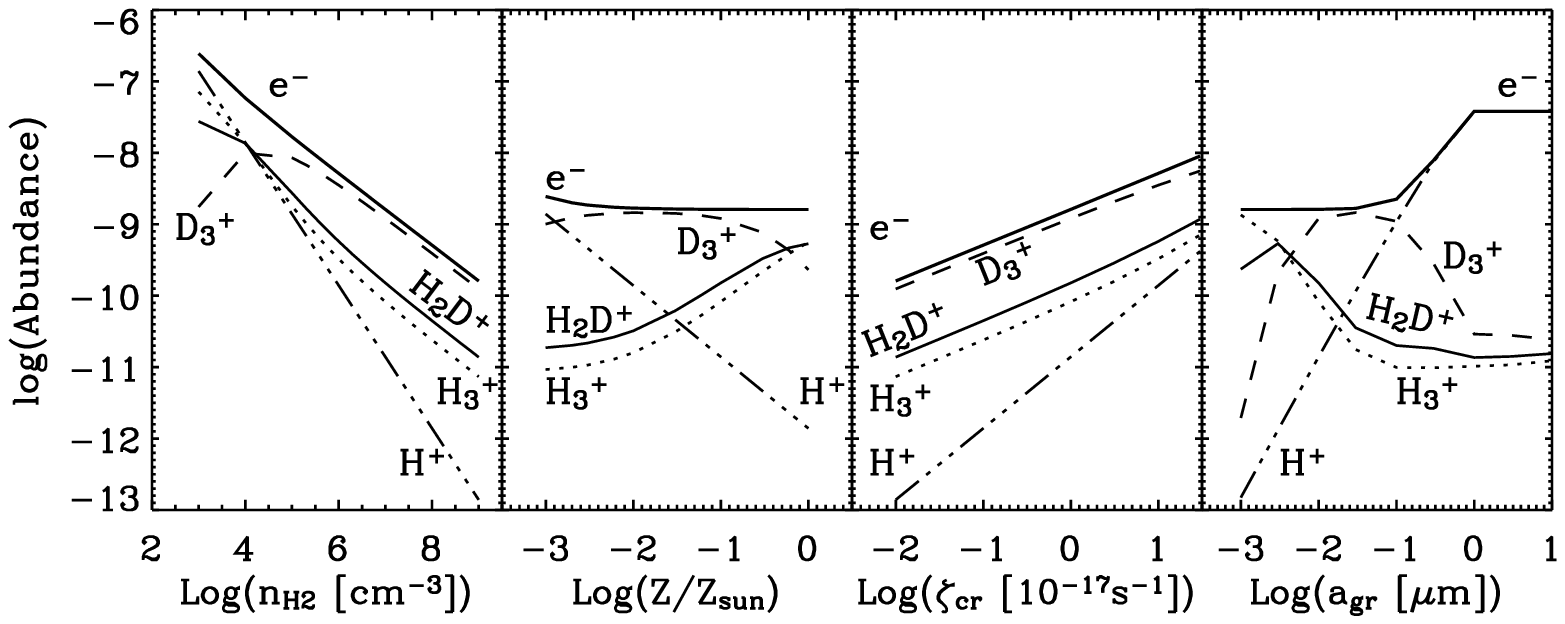}}
\caption{\label{fig:abundance}Predicted abundances of the main charge
  carriers in the dark halo cloudlets (based on the Ceccarelli \&
  Dominik (2005) model).  Form left to right, the abundances are shown
  as a function of the gas density, metallicity, cosmic ray ionization
  rate and grain radius. When varying the metallicity, we reduced the
  standard abundances of dust grains, CO and N$_2$ proportionally to
  the metallicity, relative to values typical in galactic molecular
  clouds: $x(\mathrm{CO})=9.5\times10^{-5}$,
  $x(\mathrm{N_2})=2\times10^{-5}$, and a dust-to-gas mass-ratio of
  0.01. The plots have been done for a temperature of 8 K, density
  $10^7$ cm$^{-3}$, metallicity 0.1, cosmic ray ionization rate
  $10^{-17}$ s$^{-1}$, and grain radius of 0.01 $\mu$m when these
  parameters are not varied in the relevant panel. The age of the
  clouds has been assumed larger than $\sim10^3$ yr, which is the
  timescale for the CO and N$_2$ condensation onto the grains for a
  density of $10^7$ cm$^{-3}$ (the timescale scales with the inverse of the
  density).}
\end{figure*}

Figure \ref{fig:abundance} shows the chemical composition of the
cloudlets as function of four key parameters: the density in the cloud
\nHH, the metallicity $Z$, the cosmic ray ionization rate applicable in
the halo, \zcr, and the size of dust grains present in the clouds,
\agr.  The plots show that H$_2$D$^+$ is almost never the main
positive charge carrier in the cloudlets, even though it is almost
always more abundant than H$_3^+$.  The conditions are so
extreme in the cloudlets that either D$_3^+$ or H$^+$ are the most
abundant ions.  As these are not observable, H$_2$D$^+$ remains the
best tracer.  It is very instructive
to study the dependence of the H$_2$D$^+$ abundance on the different
parameters.  First of all, the abundance deacreases as the density
increases, leading to a weak dependence of the total H$_2$D$^+$ column
density in a cloud on the cloudlet density.  This also means that the
internal structure of the cloudlets only weakly influences the
resulting column densities.  Also the dependence on the metallicity of
the clouds is shallow.  Changing it from 10$^{-3}$ times solar to
solar increases the abundance of H$_2$D$^+$ by just a factor of
30.  The abundance of all ions scales approximately with cosmic
ionization rate $\sqrt{\zcr}$.  Finally, the size of grains plays an
important role.  The H$_2$D$^+$ abundance is highest for the smallest
grain sizes.  As the grains get bigger, the H$_2$D$^+$ abundance decreases
but levels out at grain sizes of 0.1 $\mu$m.  It should be kept in mind
that the dependences discussed here are all relative to the chosen set
of parameters.  In reality, the parameter space is four dimensional
and the dependencies are complex.  However, we can derive a range of
abundances to be expected in the cloudlets.  The H$_2$D$^+$ abundance
is between 10$^{-7}$ and 10$^{-12}$, when taking into account a large
range for the four parameters varied simultaneously (larger than what
is shown in Fig. \ref{fig:abundance}).

The emission from the ground state transition of ortho-H$_2$D$^+$ at
372 GHz can be used to observationally probe cold halo cloudlets.
Like any other emission line, the ortho-H$_2$D$^+$ line intensity
depends on the excitation conditions (gas temperature and density),
and on the average column density of ortho-H$_2$D$^+$ in the telescope
beam.  Since the critical density of the ortho-H$_2$D$^+$ ground
transition is $\sim10^6$ cm$^{-3}$, the line is very likely always
thermally populated. In addition, the cloudlets temperature is likely
lower than 10 K but larger than $\sim5$ K (the condensation
temperature for H$_2$), so that the intensity very weakly (less than a
factor 4) depends on the temperature. Therefore, the most important
parameter in the prediction of the H$_2$D$^+$ line intensity is the
ortho-H$_2$D$^+$ column density, which depends on the ortho-to-para
ratio and the total H$_2$D$^+$ column density.

{\it a) H$_2$D$^+$ ortho-to-para ratio:} The chemical model gives the
total abundance of H$_2$D$^+$.  Theoretical
estimates predict that for large ($\geq 10^6$ cm$^{-3}$) densities
this ratio depends only on the gas temperature and reaches the unity
at 9 K \cite{2004A&A...427..887F}.  However, these theoretical
estimates strongly depend on the ortho-to-para ratio of H$_2$, itself
a very poorly known (=observed) parameter.  For the purpose of
estimating the detectability of the ortho-H$_2$D$^+$ line at 372 GHz,
we will consider an ortho-to-para ratio about unity.

{\it b) H$_2$D$^+$ column density:} We assume that the halo contains a mass
$\MH$ in a sphere with radius $\RH$.  The average H$_2$ column density
through this halo can be estimated to be $\NHH=\MH/\pi\RH^2\mu m_p$
where $\mu=2.8$ is the mean molecular weight in H$_2$ gas and $m_p$ is
the mass of a proton.  Note that this column is an average of lines of
sight passing through a cloudlet, and other lines of sight not passing
through one.  In an uniformly distributed halo, the average column
density of H$_2$D$^+$ is then simply given by
\begin{equation}
\label{eq:3}
\NHHDH = \frac{\MH}{\pi\mu m_\mathrm{p}\RH^2} \xHHD(\nHH,Z,\zcr,\agr)
\end{equation}
where \xHHD is the abundance of the H$_2$D$^+$ molecule relative to
H$_2$.  We will use the example of NGC3198 to have a guideline for the
choice of general halo properties.  This galaxy is considered the
prototype for studies of halo dark matter.  Interpretation of the HI
rotational curve indicates that the halo extends more than 30 kpc, and
the mass contained within 30 kpc is $1.5\times 10^{11}$ M$_\odot$
\citep{1985ApJ...295..305V}.  If this mass is present as cold H$_2$ clouds,
it corresponds to an average total column $\NHH=
\mathrm{3\times10^{21}\,cm^{-2}}$.  Using the results of the chemical
modeling above, we arrive at an estimated column density of H$_2$D$^+$
between $3\times10^{9}$ and $3\times10^{13}$ cm$^{-2}$.

{\it c) ortho-H$_2$D$^+$ line emission:} If the ortho-H$_2$D$^+$ line
is optically thin for an individual cloudlet, the line emission is
simply proportional to the column density given above. If the line is
optically thick, the emission is proportional to the column density at
which the line becomes optically thick multiplied by the telescope
filling factor.  To have an order of magnitude, the ortho-H$_2$D$^+$
line becomes optically thick at an ortho-H$_2$D$^+$ column density of
$\sim 10^{13}$, if the linewidth is $\sim$1 km/s. Therefore, the line
emission depends on the properties of individual cloudlets, which are
very different depending on the particular model used (see \S 1).  In
the following, we will consider two cases.

The first one assumes a cloudlet mass of $1M_{\odot}$ and a radius of
1000 AU \citep{1996ApJ...472...34G}, corresponding to $n_{H_2}\sim
6\times10^{7}$ cm$^{-3}$ and an average cloud $N_{H_2} \sim
1.3\times10^{24}$ cm$^{-2}$.  Using a line width of 2 km/s (twice the
virial velocity of a single cloud), an abundance of
x(H$_2$D$^+$)$\sim 3\times 10^{-11}$ is the limit for optically thin
emission. For the case of NGC3198, the filling factor of such clouds
would be $4\times10^{-3}$ when we assume a constant density of
cloudlets throughout the halo.  Thus, the maximum average
ortho-H$_2$D$^+$ column density of the halo is $\sim10^{11}$
cm$^{-2}$. For the second case, the least favorable to a H$_2$D$^+$
detection, we use the unidentified EGRET sources which
should have masses of 10$^{-3}$ M$_{\odot}$ and radii of 10 AU.  In
this case, the cloudlet $n_{H_2}$ reaches $6\times10^{10}$
cm$^{-3}$ and the $N_{H_2}$ of such a cloudlet would be
$2\times10^{25}$ cm$^{-2}$.  The ortho-H$_2$D$^+$ line would be
optically thick within the cloudlet for
$x(\mathrm{H_2D^+})>1.5\times10^{-12}$ (assuming that the linewidth is
dominated by a 0.5 km/s turbulence). For an uniform dark halo mass and
size like in NGC3198, this case corresponds to a filling factor of
$4\times10^{-4}$, and, therefore, a maximum average ortho-H$_2$D$^+$
column density of $\sim10^{10}$ cm$^{-2}$.

Note that a centrally condensed distribution of the cloudlets could
increase the filling factor and/or the average H$_2$D$^+$ column
density close to the galaxy center significantly in both cases. For
example, if the cloudlets follow a $r^{-1}$ or $r^{-2}$ power law the
average column density in a 20$''$ beam would be larger by a factor 4
and 60 respectively. In summary, the beam-averaged ortho-H$_2$D$^+$
column density in the cloudlets halo of NGC3198, if uniform, is
predicted to range between $\sim10^{9}$ and $\sim10^{13}$
cm$^{-2}$. If the halo is centrally condensed this number
increases. For some cloudlet parameters used in the literature the
emission will be optically thin, for others it will be optically
thick. The two cases discussed above would suggest a ortho-H$_2$D$^+$
column density around to $\sim10^{11}$ cm$^{-2}$ or less, if the halo
is distributed uniformly. If a $r^{-2}$ power law applies, the
ortho-H$_2$D$^+$ column density could be as high as $\sim 4\times
10^{12}$ cm$^{-2}$.

At present, the rest frequency of ortho-H$_2$D$^+$ at 372 GHz is
observable with the CSO and APEX telescopes, and, for more distant
galaxies, with JCMT, KOSMA and SMA. The line detectability depends
somewhat on the telescope employed, but, based on previous published
observations, it is safe to say that present facilities allow
detections of ortho-H$_2$D$^+$ column densities of about
$1\times10^{12}$ cm$^{-2}$
\citep{2003A&A...403L..37C,2004ApJ...607L..51C}.  Much deeper
integrations would lower this limit by about a factor 3, bringing the
detectable ortho-H$_2$D$^+$ column density to $3\times10^{11}$
cm$^{-2}$.  Therefore, we predict that the ortho-H$_2$D$^+$ 372 GHz
line should be detectable if the high end estimate of the H$_2$D$^+$
abundance in cloudlets applies.  Indeed, a positive detection of this
line would not only definitively prove the presence of important
amounts of cold molecular material in galaxies, but would also
strongly constrain the location and nature of the cloudlets.

\begin{acknowledgements}
  We warmly thank Francoise Combes for her frank and helpful
  discussions, and Dr. Scalo and an anonymous referee for comments
  which helped to clarify the Letter content. We acknowledge Travel
  support through the Dutch/French van Gogh program, project VGP
  78-387.
\end{acknowledgements}

\clearpage

%%%%%%%%%%%%%%%%%%%%%%%%%%%%%%%%%%%%%%%%%%%%%%%%%%%%%%%%%%%%%%%%%%%%%%%%%%%%
\bibliographystyle{apj}
\bibliography{/Users/cecilia/common/ceccarellic}

%%%%%%%%%%%%%%%%%%%%%%%%%%%%%%%%%%%%%%%%%%%%%%%%%%%%%%%%%%%%%%%%%%%%%%%%%%%%
\clearpage

\end{document}